\documentclass[aps,prl,floats,twocolumn,floatfix,showpacs]{revtex4}%
\usepackage{psfig,epsf,graphicx}
\usepackage{amssymb}
\usepackage{amsmath}
\usepackage{eepic}
\usepackage{amsfonts}
\usepackage[psamsfonts]{eucal}
\usepackage{eucal}
\usepackage{graphicx}%
\newlength{\piclen}
\begin{document}

\title{Pressure-induced metal-insulator transition in LaMnO$_{3}$ is not of
Mott-Hubbard type}
\author{A.\ Yamasaki, M.\ Feldbacher, Y.-F.\ Yang, O.\ K.\ Andersen, and K.\ Held}
\affiliation{Max-Planck Institut f\"{u}r Festk\"{o}rperforschung, D-70569 Stuttgart, Germany}
\date{\today}

\begin{abstract}
Calculations employing the local density approximation combined with static
and dynamical mean-field theories (LDA+$U$ and LDA+DMFT) indicate that the
metal-insulator transition observed at 32\thinspace GPa in paramagnetic
LaMnO$_{3}$ at room temperature is not a Mott-Hubbard transition, but is
caused by orbital splitting of the majority-spin $e_{g}$ bands. For
LaMnO$_{3}$ to be insulating at pressures below 32\thinspace GPa, both on-site
Coulomb repulsion and Jahn-Teller distortion are needed.
\end{abstract}
\maketitle

Since the discovery of colossal magnetoresistance (CMR) \cite{CMR}, manganites
have been intensively studied. The key to understand CMR is the
high-temperature paramagnetic insulating-like phase, which is characterized
not only by an increase of resistivity with decreasing temperature, but also
by unusual dynamical properties, such as low spectral weight at the Fermi
level for a wide range of doping \cite{PES,Park,optics}. Theoretical understanding of this hole-doped
paramagnetic phase remains incomplete, and CMR transition temperatures are
lower than technologically desirable.

In this Letter, we shall focus on the pressure-induced insulator-metal (IM)
transition in the undoped parent compound LaMnO$_{3}$ with configuration
$t_{2g}^{3}e_{g}.$ This transition occurs at room temperature, well above the
magnetic ordering temperature $\left(  T_{N}\mathrm{=}140\,\text{K}\right)  ,$
well below the cooperative Jahn-Teller (JT) temperature 
($T_{\mathrm{oo}}\mathrm{=}%
740$\thinspace K at 0\thinspace GPa), and at a hydrostatic pressure of
32\thinspace GPa where the JT distortion appears to be completely suppressed
\cite{Loa}. The IM transition thus seems to be a bandwidth-driven Mott-Hubbard
transition of the $e_{g}$ electrons and points to the dominating importance of
the Coulomb repulsion between two $e_{g}$ electrons on the same site. This is
supported by recent self-interaction-corrected local density approximation 
(LDA) calculations, performed,
however, for the cubic structure and magnetically ordered states at low
temperature \cite{SIC}. 
Structural distortions at 0 K as functions of pressure were recently calculated
with the LDA+$U$ method~\cite{Trimarchi}.
On the theoretical side, it has been an issue of long
debate whether the JT distortion or the Coulomb repulsion is
responsible for the insulating behavior of LaMnO$_{3}$ at normal pressure. The
high-pressure experiment \cite{Loa} seems to favor the latter.

Here, we shall study the room-temperature electronic structure of LaMnO$_{3}$
at normal pressure and the pressure-induced IM transition by means of LDA + $U$
\cite{LDAU} and LDA + dynamical mean field theory (DMFT) \cite{LDADMFT} 
calculations. Upon going from the
insulating to the metallic, high-pressure phase, we shall find that the
orbital polarization and the concomitant splitting of the two majority-spin
$e_{g}$ bands are gradually reduced. The IM transition takes place when the
bands start to overlap. Since this occurs within the 
(orbitally) symmetry-broken phase,
this IM transition is not a Mott-Hubbard transition. The Coulomb interaction,
\emph{as well as} the JT distortion are needed for a proper
description of this transition and the insulating nature of LaMnO$_{3}$.

\begin{figure}[b]
\includegraphics[width=0.9\columnwidth]{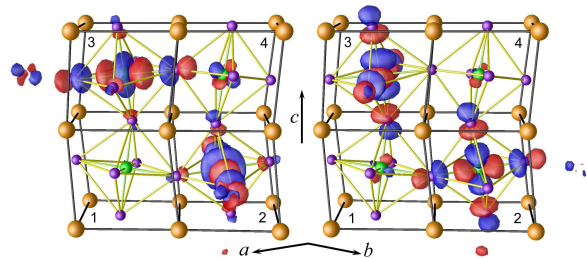} 
\caption{(color). LaMnO$_{3}%
$ orthorhombic translational cell (\textit{Pbnm}) and LDA-NMTO Mn $e_{g}$
crystal-field orbitals $\left\vert 1\right\rangle $ (left) and $\left\vert
2\right\rangle $ (right) of, respectively, lowest and highest energy. For the
sake of clarity, the orbitals have been placed only in subcells 3 and 2; those
in subcells 1 and 4 may be obtained by the LaO mirror plane perpendicular to
the $c\,(\mathrm{=}z)$ axis. Since they have antibonding O\thinspace2$p$
tails, orbitals $\left\vert 1\right\rangle $ and $\left\vert 2\right\rangle $
are directed, respectively, along and perpendicular to the longest Mn-O bond.
Red/blue indicates a positive/negative sign.}%
\label{Fig:Orb}%
%\vspace{-.0cm}
\end{figure}

The orthorhombic crystal structure of LaMnO$_{3}$ at atmospheric pressure is
shown in Fig.\thinspace\ref{Fig:Orb}. The O$_{6}$ octahedra are elongated in
the $y$ direction (nearly parallel to $\mathbf{b}-\mathbf{a}$ in Fig.\thinspace\ref{Fig:Orb}) in subcells 1
and 3, and in the $x$ direction (nearly parallel to $\mathbf{b}+\mathbf{a})$
in subcells 2 and 4. This JT distortion decreases linearly from 11\% at
0\thinspace GPa, to 4\% at 11\thinspace GPa, the highest pressure for which
internal parameters were measured \cite{Loa}. The GdFeO$_{3}$-type distortion
tilts the corner sharing octahedra around the $b$ axis and rotates them around
the $c$ axis, both in alternating directions. When the pressure increases from
0 to 11 GPa, the tilt is reduced from 12$%
{{}^\circ}%
$ to 8$%
{{}^\circ}%
$ and the rotation from 7$%
{{}^\circ}%
$ to 5$%
{{}^\circ}%
$.

\begin{figure}[tb]
\includegraphics[width=1.0\columnwidth]{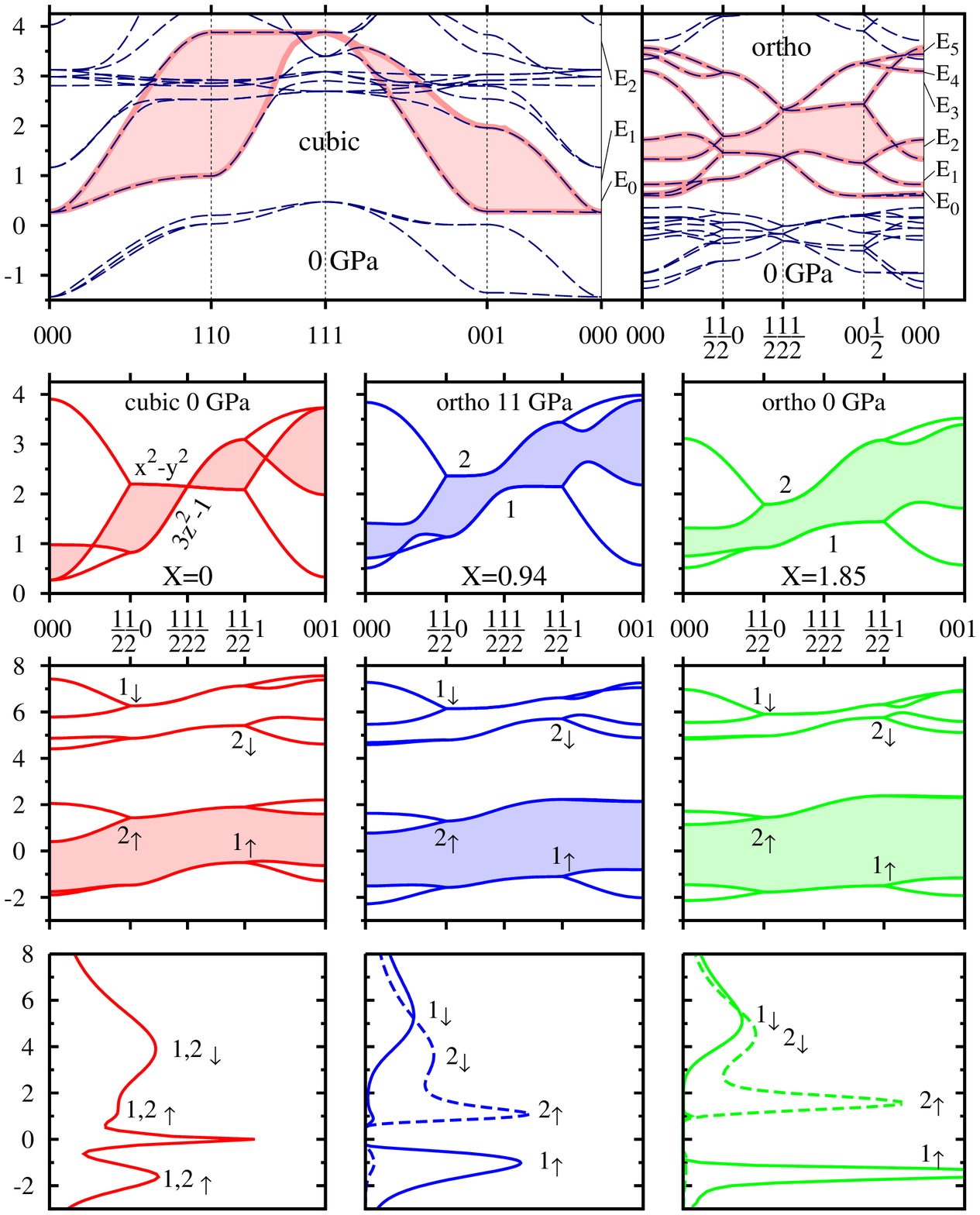}
\caption{(color).
\textit{Top:} Paramagnetic LDA band structure for orthorhombic (right) and
hypothetical cubic (left) LaMnO$_{3}$ at 0\thinspace GPa plotted along the
high-symmetry lines in the $k_{x}\mathrm{=}k_{y}$ plane. Energies are in eV,
the $\mathbf{k}$ unit is $\pi$ \cite{NJP}, and the $\mathbf{k}$ points
marked are $\Gamma$MRX$\Gamma$ in the cubic, and $\Gamma$YTZ$\Gamma$ in the
orthorhombic BZ. The latter is folded in from the former and
has the following smallest inequivalent reciprocal-lattice vectors:
$\mathbf{Q}=000,$ $110,$ $111,$ and $001$. Dashed blue bands: large NMTO basis
set; red bands: Mn $e_{g}$ NMTO basis set employed in Eq.\thinspace(\ref{H}).
The N+1 support energies $E_{i}$ are shown at the right-hand sides. The zero
of energy corresponds to configuration $t_{2g}^{4}.$ \textit{Second row}: 0
and 11\thinspace GPa orthorhombic (0\thinspace GPa cubic) $e_{g}$ bands folded
out (in) to the $\left(  000,110\right)  $-BZ. The dimensionless band-shape
parameter, $X,$ is the LDA crystal-field splitting in units of the effective
hopping integral $t\mathrm{\equiv}\left\vert t_{dd\sigma}\right\vert
\mathrm{\sim}W/6,$ both obtained from the NMTO Mn $e_{g}$ Wannier functions.
\textit{Third row}: As second row, but obtained by spin-polarized LDA+$U$ for
random spin orientations (room temperature). Bands are labelled by their main
character. The zero of energy is the Fermi level. \textit{Bottom}: Spectra
calculated by LDA+DMFT. The full and dashed lines give the projections onto
orbitals $\left\vert 1\right\rangle $ and $\left\vert 2\right\rangle $,
respectively.}%
\label{Fig:Bands}%
%\vspace{-.4cm}
\end{figure}

The top right-hand side of Fig.\thinspace\ref{Fig:Bands} shows the
paramagnetic LDA bands at normal pressure and the left-hand side those for a
hypothetical, cubic structure with the same volume. The dashed blue bands were
obtained with a large basis set of $N$th-order muffin tin orbitals (NMTOs)
\cite{NMTO,NJP}. Within the frame of the figure, we see the 3$\left(
\times4\right)  $ Mn\thinspace$t_{2g}$ and the 2$\left(  \times4\right)  $
Mn\thinspace$e_{g}$ bands, while the O\thinspace$2p$ bands are below. As a
check, we also carried out spin-polarized calculations for ferro- and
antiferromagnetic $A$-type orders and found full agreement with previous work
\cite{LDAcalcs}. Near the top of Fig.\thinspace\ref{Fig:Bands} ---and
continuing above it--- is the La\thinspace$5d$ band which is pushed 2\thinspace
eV up by $pd\sigma$ hybridization with oxygen when going from the cubic to the
orthorhombic structure. This hybridization is a reason for the GdFeO$_{3}%
$-type distortion \cite{NJP}, but since it only involves O\thinspace$p$
orbitals perpendicular to the one which hybridizes with Mn\thinspace$e_{g}$
(Fig.\thinspace\ref{Fig:Orb}), the La\thinspace5$d$ and Mn\thinspace$e_{g}$
bands hardly hybridize. Finally, the narrow band crossing the cubic
Mn\thinspace$e_{g}$ band is La\thinspace4$f.$

The NMTO method can be used to generate minimal basis sets \cite{NMTO,NJP},
such as the Mn\thinspace$e_{g}$ basis whose orbitals are shown in
Fig.\thinspace\ref{Fig:Orb}. This basis gives rise to the red solid bands in
the topmost panel of Fig.\thinspace\ref{Fig:Bands}, which are seen to follow
the blue dashed bands exactly, except where the latter have avoided crossings
with La bands. When their energy mesh is converged, the symmetrically
orthonormalized, minimal NMTO set is a set of Wannier functions. The $e_{g}$
NMTOs are localized by the requirement that a Mn $e_{g}$ orbital has \emph{no}
$e_{g}$ character on neighboring Mn atoms, and this confines the NMTO-Wannier
functions to being essentially as localized as those in  \cite{MaxLoc}.

Taking the Coulomb repulsion and Hund's exchange into account, three electrons
localize in the $t_{2g}$ orbitals which we describe in the following by a
classical spin $S$. These $t_{2g}$ spins, which we assume to have random
orientations at room temperature, couple to the $e_{g}$ electrons with
strength $2\mathcal{J}S\mathrm{=}2.7$\thinspace eV, as estimated by the
splitting of the $e_{g}^{\uparrow}$ and $e_{g}^{\downarrow}$ bands in our
ferromagnetic NMTO calculation (not shown). This results in the following
low-energy Hamiltonian for the two $e_{g}$-bands:
\begin{align}
\!\!\hat{H} & \!\!=\!\!\!\!\sum_{ijmn\sigma\sigma^{\prime}}%
h_{im,jn}\,\hat{c}_{im\sigma}^{\dagger}u_{\sigma\sigma^{\prime}}^{ij}\hat
{c}_{jn\sigma^{\prime}}-\mathcal{J}S\sum_{im}\left(  \hat{n}_{im\uparrow}%
\!-\!\hat{n}_{im\downarrow}\right)  \nonumber\\
&  \!+U\sum_{im}\hat{n}_{im\uparrow}\hat{n}_{im\downarrow}+\sum_{i\,\sigma
\sigma^{\prime}}(U^{\prime}-\delta_{\sigma\sigma^{\prime}}J)\hat{n}_{i1\sigma
}\hat{n}_{i2\sigma^{\prime}}\,,\label{H}%
\end{align}
where $h_{im,jn}$ is the LDA Hamiltonian in the representation of the two
($m\mathrm{=}1,2$) $e_{g}$ NMTO-Wannier orbitals per site (Fig.\thinspace
\ref{Fig:Orb} and two top rows of Fig.\ref{Fig:Bands}); $u_{\sigma
\sigma^{\prime}}^{ij}$ accounts for the rotation of the spin quantization axis
(parallel to the $t_{2g}$ spin) from Mn sites $j$ to $i$. The second line
describes the Coulomb interactions between $e_{g}$ electrons in the same ($U$)
and in different orbitals ($U^{\prime}$); $J$ is the $e_{g}$-$e_{g}$
Hund's rule exchange. We take $U\mathrm{=}5\,$eV and $J\mathrm{=}0.75\,$eV
from the literature \cite{Park}; by symmetry, $U^{\prime}=U-2J$. These values
are reasonable, also in comparison with those used for other transition-metal
oxides. Whereas a larger $U$ is appropriate when all five $d$ degrees of
freedom are treated \cite{LDAcalcs}, the smaller value used for our $e_{g}$
Hamiltonian takes the screening by the $t_{2g}$ electrons into account.

We first solve (\ref{H}) by DMFT \cite{DMFT} using quantum Monte Carlo (QMC)
simulations at room temperature. Previous calculations for LaMnO$_{3}$ with
electronic correlations, but simplified hopping integrals, include
Ref.\thinspace\cite{Imada98a}. We neglected the (orbital) off-diagonal
elements of the on-site Green function, forcing the DMFT density to have the
same symmetry as the LDA crystal field (Fig.\thinspace\ref{Fig:Orb}), a good
approximation for $e_{g}$ systems.

The spectral densities calculated using the observed crystal structures at 0
and 11\thinspace GPa \cite{Loa} are shown at the bottom of
Fig.\textbf{\thinspace}\ref{Fig:Bands} (green and blue, respectively). At
0\thinspace GPa, we find strong orbital polarization, a 2\thinspace eV gap,
and, above 3\thinspace eV, spectral densities which correspond to $e_{g1}%
^{\downarrow}$ and $e_{g2}^{\downarrow}$ configurations antiparallel to the
$t_{2g}$ spin at that site. Experiments for undoped LaMnO$_{3}$ show similar
---albeit less sharp--- gaps \cite{PES,Park,optics}. 
At 11\thinspace GPa, we find
the gap to be halved, but the orbital polarization to be just slightly reduced.

\begin{figure}[b]
\rotatebox{270}{\includegraphics[height=0.95\columnwidth]{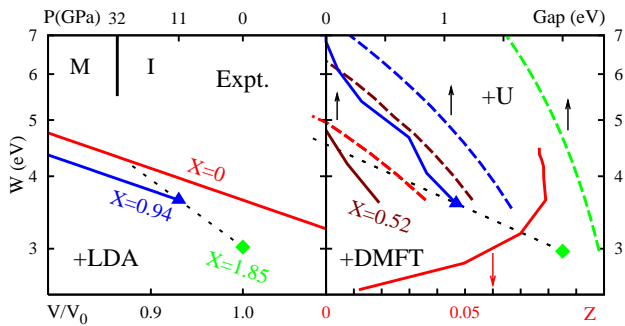}}%
\caption{(color).
\textit{Left}: LDA $e_{g}$ bandwidths, $W\mathrm{\sim}6t,$ calculated
as functions of compression, $V/V_{0}.$ The top abscissa gives the
experimental pressures, with 32\thinspace GPa marking the observed IM
transition \cite{Loa}. \textit{Right}: 300\thinspace K LDA+DMFT (solid lines)
and LDA+$U$ (dashed lines) results calculated as functions of $W.$ For
insulators, we plot the DOS gap (top scale) and for metals the quasiparticle
weights (bottom scale). Full symbols indicate actual experimental structures
and connecting black dotted lines are extrapolations. Each curve was
calculated with fixed structure type (band shape): orthorhombic 0\thinspace
GPa (green, $X\mathrm{=}1.85$), orthorhombic 11\thinspace GPa (blue,
$X\mathrm{=}0.94$), cubic plus crystal-field splitting (dark red, $X\mathrm{=}0.52$),
and cubic (red, $X\mathrm{=}0$).\label{Fig:MIT}}%
%\vspace{-.3cm}
\end{figure}

Since for higher pressures the internal structural parameters are unknown, we
scaled the 11\thinspace GPa structure uniformly and calculated the LDA
$e_{g}$ bandwidth $W$ as a function of compression. The result is shown in
blue on the left-hand side of Fig.\thinspace\ref{Fig:MIT}. On the right-hand
side, we show the gap obtained from LDA+DMFT for various values of $W$ (blue
solid line), keeping the LDA band shape constant. We see that the gap
decreases, but does not close before $W$=7\thinspace eV, which corresponds to
a pressure way beyond the experimental 32\thinspace GPa.

Let us now turn to the cubic phase (red), for which LDA+DMFT yields
\emph{metallic} behavior already at 0\textbf{\thinspace}GPa (left-hand side
of Fig.\textbf{\thinspace}\ref{Fig:Bands}) \cite{Note}. The spectral density
exhibits a quasiparticle peak at $\varepsilon_{F}$, lower and upper Hubbard
bands at about $\mp1.5\,$eV, and a peak due to the antiparallel-spin
configurations around $4\,$eV. In the right-hand side of Fig.\thinspace
\ref{Fig:MIT}, we show how the quasiparticle weight $Z$ decreases with volume
expansion, and does not vanish before $W$=2.7\thinspace eV. One needs to
\emph{pull} the cubic crystal to make it insulating! Both types of structure,
the cubic ($X$=0, red) and the one observed at 11\thinspace GPa ($X$=0.94,
blue) give grossly wrong pressures for the IM transition ---way beyond the
expected errors of the LDA and DMFT approximations. Hence, we can conclude
that the IM transition occurs at a reduced (compared to the one at 11GPa), but
finite distortion \cite{Note}. Such an intermediate distortion of
$X\mathrm{=}0.52$ (dark red line) indeed yields the IM transition at roughly
the correct experimental pressure as seen by using the left hand-side of
Fig.\thinspace\ref{Fig:MIT} to translate bandwidths into pressures. A finite
distortion at the IM transition is also in accord with the observed noncubic
crystal structure at the IM transition \cite{Loa}. Besides the JT
distortion, the Coulomb interaction is absolutely essential to describe
insulating LaMnO$_{3}$. Without it, one would just obtain the metallic LDA
bands shown in the second row of Fig.\thinspace\ref{Fig:Bands}, albeit with a width reduced
by $\frac{2}{3}$ due to the random spin orientation.

A detaileded understanding of the band structure in the insulating phase may
be obtained by treating the $e_{g}$ Hamiltonian (\ref{H}) in the static mean
field approximation (LDA+$U$) as done by Ahn and Millis \cite{Millis}, but
using the accurate NMTO Hamiltonian. This is supported by the similarity of
the resulting LDA+$U$ gap-versus-$W$ 
curves shown by the dashed green, blue, dark
red, and red lines in the right-hand side of Fig.\thinspace\ref{Fig:MIT} to
those obtained by DMFT. The LDA+$U$ treatment overestimates the gaps~\cite{giorgio}
, but
since it is easy to follow, we use it in the remainder of this Letter to
explain ---almost analytically--- the development of the electronic structure,
starting with the cubic $e_{g}$ LDA bands, then turning on the orthorhombic
distortions, and finally the spin and the $e_{g}$-$t_{2g}$ and $e_{g}$-$e_{g}$
Coulomb interactions.

In the cubic structure, the two $e_{g}$ bands never cross, as seen in the top
left-hand side of Fig.\thinspace\ref{Fig:Bands}. Since the dominating,
indirect $dpd,$ as well as the direct $dd$ contributions to the hopping
integrals (off-site elements of $h_{im,jn})$ between nearest neighbors in the
$z\mathrm{=}0$ plane are bonding (negative), the bottom of the band is at 000
and the top at 111$.$ In the $k_{x}\mathrm{=}k_{y}$ plane the eigenfunctions
are the $d_{3z^{2}-1}\left(  \mathbf{k}\right)  $ and $d_{x^{2}-y^{2}}\left(
\mathbf{k}\right)  $ Bloch functions, with the former band lying lower. The
hopping integrals are close to those of a first-nearest-neighbor tight-binding
(TB) model with $t_{dd\delta}\mathrm{=}0$ and $t_{dd\sigma}\mathrm{\equiv}-t.$
In this model, the cubic bandwidth is $W\mathrm{=}6t,$ and the band is
symmetric around its center: $\varepsilon_{2}\left(  \mathbf{k}\right)
\!=\!-\varepsilon_{1}\left(  \mathbf{k-}\left[  111\right]  \pi\right)$. 
In the
$k_{x}\mathrm{=}k_{y}\mathrm{\equiv}k$ plane, the dispersions of the
$d_{3z^{2}-1}$ and $d_{x^{2}-y^{2}}$ bands are,
respectively, $-\!t\cos k\!-\!2t\cos k_{z}$ and $-\!3t\cos k.$

In order to understand the orthorhombic bands in the top right-hand side of
Fig.\thinspace\ref{Fig:Bands}, one should first note that mirroring an $e_{g}$
orbital in the LaO plane is equivalent to translating it along $z.$ The 8
orthorhombic bands can therefore be folded \emph{out} across the 
Brillouin zone (BZ) face
$k_{z}\mathrm{=}\frac{\pi}{2}$ to the 4 green bands shown in the second row.
The blue bands were calculated for the less distorted 11\thinspace GPa
structure, and the 4 red bands are the 2 cubic bands, folded \emph{into} the
$\mathbf{Q}\mathrm{=}\left(  110\right)  \pi$ BZ \cite{NJP}. We see that
the dominating effect of the distortion is to couple the bands, and thereby to
\emph{cut} the $e_{g}$ band \emph{in two, }which overlap by \emph{less than}
$W$. For $k_{x}\mathrm{=}k_{y},$ there is no coupling between $\varepsilon
_{x^{2}-y^{2}}\left(  \mathbf{k}\right)  $ and $\varepsilon_{3z^{2}-1}\left(
\mathbf{k}\right)  ,$ but only between $\varepsilon_{3z^{2}-1}\left(
\mathbf{k}\right)  $ and $\varepsilon_{x^{2}-y^{2}}\left(  \mathbf{k}%
-\mathbf{Q}\right)  $. This allows us to fold the 4 bands out to merely 2,
letting not only $k_{z},$ but also $k_{x}\mathrm{=}k_{y}\mathrm{\equiv}k$ run
from $-\pi$ to $\pi$. In the TB model, these two bands are the eigenvalues of
a $2\mathrm{\times}2$ matrix with diagonal elements $-t\cos k\!-\!2t\cos
k_{z}\!+\!\left(  \Delta\!-\!\sqrt{3}\delta\right)  /4$ 
and $3t\cos k\!-\!\left(
\Delta\!-\!\sqrt{3}\delta\right)  /4$ between, respectively, 
$d_{3z^{2}-1}\left(
\mathbf{k}\right)  $ and $d_{x^{2}-y^{2}}\left(  \mathbf{k-Q}\right)  $, and
with the off-diagonal element $-\left(  \sqrt{3}\Delta+\delta\right)  /4.$
Here, $\mp\Delta/2$ are the energies of the $d_{3y^{2}-1}$ and $d_{x^{2}%
-z^{2}}$ orbitals in cell 3 (Fig.\thinspace\ref{Fig:Orb}), $-\delta/2$ is the
matrix element between them, and $\left(  \Delta^{2}\!+\!\delta^{2}\right)
^{1/2}\mathrm{\equiv}Xt$ is the crystal-field splitting. We see that the
$e_{g}$ band is cut $\frac{3}{2}t\mathrm{=}\frac{1}{4}W$ above the bottom and
$\frac{3}{2}t$ below the top, with gaps of size $Xt$, so that the band
\emph{overlap} is merely $\left(  3\!-\!X\right)  t\!+\!\mathcal{O}\left(
X^{2}\right)  .$ For the 11 and 0\thinspace GPa structures, respectively,
$\left(  \Delta,\delta\right)  $=$\left(  500,110\right)  $ and $\left(
849,126\right)  $\thinspace meV, and $X$=0.94 and $1.85$. For large $X,$ the
density of states (DOS) \emph{gap} is found to be $\left(  X\!-\!4\right)
t\!+\!\mathcal{O}\left(  X^{-2}\right)  .$

Inclusion of the spin and the $e_{g}$-$t_{2g}$ repulsion in the static
mean field approximation splits the $e_{g}$ band by $2\mathcal{J}S$ into
$e_{g}^{\uparrow}$ and $e_{g}^{\downarrow}$ bands with spins locally parallel
and antiparallel to that of the $t_{2g}$ spin. 
Because of the random orientation
of the latter, the $e_{g}$ hopping integrals are independent of spin and
\emph{reduced} by the factor $\frac{2}{3}$ \cite{Millis}. Since the coupling
between the $e_{g}^{\uparrow}$ and $e_{g}^{\downarrow}$ bands is of order
$t^{\prime}{}^{2}/2\mathcal{J}S,$ the DOS gap remains $\left(  X^{\prime
}-3.6\right)  t^{\prime}$ to order $t^{\prime}.$ Here, $t^{\prime
}\mathrm{\equiv}\frac{2}{3}t,$ $X^{\prime}t^{\prime}\mathrm{\equiv}Xt$ is the
crystal-field splitting, and the value 3.6 for the IM transition was obtained
numerically; it lies between the small and large-$X$ limits.

Finally, the $e_{g1}^{\uparrow}$-$e_{g2}^{\uparrow}$ Coulomb repulsion,
$U^{\prime\prime}\approx U^{\prime}\!-\!J\!=\!2.75$\thinspace eV, splits the occupied
and empty $e_{g}^{\uparrow}$ bands apart by the \emph{effective} crystal
field:%
\begin{equation}
\left(  \Delta^{2}+\delta^{2}\right)  _{eff}^{{1}/{2}}\approx\left(
\Delta^{2}+\delta^{2}\right)  ^{1/2}+\left(  n_{1}^{\uparrow}-n_{2}^{\uparrow
}\right)  U^{\prime\prime}.\label{sc}%
\end{equation}
The orbital polarization, $n_{1}^{\uparrow}-n_{2}^{\uparrow}\equiv P\left(
X_{eff}^{\prime}\right)  ,$ is a band-structure function which increases
linearly from $0,$ reaches $0.8$ for $X_{eff}^{\prime}\mathrm{=}3.6,$ and
saturates at 1 for large $X_{eff}^{\prime}.$ As a consequence, Eq.\thinspace
(\ref{sc}) written in the form: $\frac{2}{3}X_{eff}^{\prime}\!=\!X\!+\!P(X_{eff}%
^{\prime})U^{\prime\prime}/t$ is the self-consistency condition for the
effective crystal-field splitting and orbital polarization as functions of the
band-structure parameters $t$ and $X.$ This equation explains the dashed lines
in Fig.\thinspace\ref{Fig:MIT} and yields the band structures shown in the
third line of Fig.\thinspace\ref{Fig:Bands}. Note that, in contrast to 
DMFT, the LDA+$U$ approximation gives a spontaneous orbital polarization of
the cubic band structure at normal volume.

In conclusion, the IM transition in LaMnO$_{3}$ at 32 GPa is not of
Mott-Hubbard type. Rather, it is triggered by small distortions which create a
crystal-field splitting, strongly enhanced by the Coulomb repulsion. For
sufficiently large splitting, the majority-spin $e_{g}$ bands separate and
LaMnO$_{3}$ becomes an insulator. Crucial are also the bandwidth reductions of
$\frac{2}{3}$ and $\frac{3.6}{6}$ arising from, respectively, the spatially
uncorrelated directions of the $t_{2g}$ spins at room temperature and the
cutting by the effective crystal field of the $e_{g}$ band in two subbands
which overlap by less than the bandwidth.

We acknowledge helpful discussions with O. Gunnarsson, P. Horsch, G.
Khaliullin, I.\ Loa, and K.\ Syassen, as well as support by the Deutsche
Forschungsgemeinschaft through the Emmy Noether program (K.H.,M.F.).

{\it Note added in proof.}---
An LDA+$U$ study similar to ours, albeit for 0 K and 0 GPa, was reported
in Ref.~\cite{Yin}.

\end{document}